\documentclass{article}

\usepackage{bm}
\usepackage{amsmath,amssymb}
\usepackage{epsfig}
\usepackage{latexsym} 
\usepackage{multicol}

\def\R{\mathbb{R}}
\def\C{\mathbb{C}}
\def\DD{{\bf D}}
\def\EE{{\bf E}}
\def\FF{{\bf F}}
\def\GG{{\bf G}}
\def\JJ{{\bf J}}

\def\cc{{\bf c}}
\def\rhodot{\dot{\rho}}
\def\rr{{\bf r}}
\def\q{{\bf q}}  
\def\dq{\dot{\bf q}} 
\def\hrho{\hat{\bm{\rho}}} 
\def\brho{\bm{\rho}} 
\def\hrhoa{\hat{\bm{\rho}}_{\alpha}} 
\def\hrhod{\hat{\bm{\rho}}_{\delta}}
\def\alphadot{\dot{\alpha}} 
\def\deltadot{\dot{\delta}} 
\def\hn{{\bf \hat{n}}}
\def\hv{{\bf \hat{v}}}
\def\n12{{\bf n_{12}}} 
\def\Att{{\cal A}}
\def\br{{\bf r}}   
\def\dbr{{\bf \dot{r}}} 
\def\dq{\dot{\bf q}} 
\def\L{{\bf L}}
\def\Rvec{{\bf R}}
\def\Rcal{\mathcal{R}}

\def\Avec{{\bf A}}

\def\bPhi{\bm{\Phi}}
\def\bPsi{\bm{\Psi}}
\def\bDelta{\Delta_{1,2}}

\def\phkstar{p_{h,k}^{\star}}
\def\phhstar{p_{h,h}^{\star}}
\def\cstar{c_\star}

\def\aq{\alpha_q}

\def\daq{\dot\alpha_q}
\def\ddaq{\ddot\alpha_q}

\def\delq{\delta_q}
\def\ddelq{\dot\delta_q}
\def\dddelq{\ddot\delta_q}

\def\bxi{{\bm\xi}}
\def\bx{{\bf x}}

\begin{document}
\title{\bf Orbit Determination with the two-body Integrals}

\author{G.~F. Gronchi,\footnote{Giovanni F. Gronchi: 
     Dipartimento di Matematica, Universit\`a di Pisa, 
     Largo B. Pontecorvo, 5, Pisa, Italy {\tt gronchi@dm.unipi.it}}
      \ L. Dimare,\footnote{Linda Dimare:
     Dipartimento di Matematica,  Universit\`a di Roma `La Sapienza',
     P.le Aldo Moro, 2, Roma, Italy
     {\tt dimare@mat.uniroma1.it}}
      \ A. Milani\footnote{Andrea Milani:
     Dipartimento di Matematica, Universit\`a di Pisa, 
     Largo B. Pontecorvo, 5, Pisa, Italy
     {\tt milani@dm.unipi.it}}
}

\date{}

\maketitle

\begin{abstract}
We investigate a method to compute a finite set of preliminary orbits
for solar system bodies using the first integrals of the Kepler
problem.  This method is thought for the applications to the modern
sets of astrometric observations, where often the information
contained in the observations allows only to compute, by interpolation,
two angular positions of the observed body and their time derivatives
at a given epoch; we call this set of data {\em attributable}.
Given two attributables of the same body at two different epochs we
can use the energy and angular momentum integrals of the two-body
problem to write a system of polynomial equations for the topocentric
distance and the radial velocity at the two epochs. We define two
different algorithms for the computation of the solutions, based on
different ways to perform elimination of variables and obtain a
univariate polynomial.
Moreover we use the redundancy of the data to test the hypothesis
that two attributables belong to the same body ({\em linkage problem}). 
It is also possible to compute a covariance matrix, describing the
uncertainty of the preliminary orbits which results from the
observation error statistics.
The performance of this method has been investigated by using a large
set of simulated observations of the Pan-STARRS project.

 
\end{abstract}

\section{Introduction}

With the new observational techniques of the next generation surveys,
like Pan-STARRS and LSST,\footnote{see the web pages {\tt
http://pan-starrs.ifa.hawaii.edu}, and {\tt http://www.lsst.org}} the
number of moving objects detected in each night of observations is
expected to increase by two orders of magnitude with respect to the
current surveys.  To deal with this huge amount of data the interest
in the study of orbit determination methods has been renewed, both
from the theoretical and the computational point of view.  The
classical methods of preliminary orbit determination by
Laplace~\cite{laplace} and Gauss~\cite{gauss}, that have been often
revisited in the last two centuries \cite{poincare}, \cite{leuschner},
\cite{merton}, are based on the knowledge of at least three
observations of a solar system body in three different nights. 
Both Laplace's and Gauss' method may produce more than one preliminary
orbit for the same object: a detailed analysis of the occurrence of
multiple solutions is in \cite{gronchi09}.
The determination of a preliminary orbit is followed by the {\em
differential corrections} \cite{bate}, an iterative method to obtain
the minimum of a target function, that improves the orbit in the sense
of the least squares fit of the residuals: this sequence of operations
was already proposed in \cite{gauss}.

The data of the current surveys generally do not provide a single
observation for an object in an observing night: in fact the
moving objects are distinguished from fixed stars by detecting them a
few times in the same night: the sequence of observations usually
gives a {\em too short arc} on the celestial sphere \cite{miletal07},
such that the data are not enough to compute an orbit for that body.
As the number of detected objects per night is very large, it is
difficult to decide whether two sequences of observations made in
different nights belong to the same object: this gives rise to the
problem of {\em linkage} of two short arcs of observations.

The information contained in a short arc of observations can be used
to define an {\em attributable}~\cite{ident4},
consisting of the angular position and velocity of the body on the
celestial sphere at a given time; the topocentric distance and the
radial velocity at that time are unknown. Therefore two short arcs of
observations belonging to the same object provide us with 8 scalar
data, from which we can try to compute an orbit.

In 1977 Taff and Hall proposed to use the angular momentum and the
energy integrals to perform orbit determination starting from a data
set that corresponds to two attributables of the same observed body
(see \cite{taff77}, \cite{taff80}).  They noticed that the problem can
be written in an algebraic form but, since the total degree is high,
they suggested to use a Newton-Raphson method to solve the problem.
This approach deals with the solutions only locally, and there are
alternative possible solutions that can be lost.
%
In this paper we shall start from the same first integrals of the
Kepler problem, but we shall exploit the algebraic character of the
problem, keeping in this way a global control on the solutions.  In
particular we shall present two different methods to solve by
elimination the polynomial system corresponding to this problem, and
to compute all the related preliminary orbits defined by the two
attributables.

Since an orbit is defined by 6 scalar data, the available information
is redundant, and we can use this to set compatibility conditions for
the solutions (see (\ref{compatible})), that should be fulfilled if
the two attributables belong to the same solar system object.
The unavoidable errors in the observations affect also the computation
of the attributables.  Given a covariance matrix for the two
attributables, expressing their uncertainty, we can use this to
compute the value of an {\em identification norm}, based on the
compatibility conditions, to decide if the attributables may be
related to the same body (i.e. if the linkage is successful) and to
choose among possible alternative solutions.

The plan of the paper is the following: after introducing some
notation related to attributables in Section~\ref{s:attrib}, we
explain in Section~\ref{s:linkage} the orbit determination method and
derive the bivariate polynomial system whose roots give us the
topocentric distances of the observed object at the two epochs.  In
Section~\ref{s:algor} we explain the two algorithms that we propose to
search for these roots and the effective computation of the orbits.
In Section~\ref{s:uncert} we deal with the uncertainty of the data,
introduce the identification norm and provide the covariance matrices
of the preliminary orbits.  Some numerical experiments are presented
in the last two sections: in Section~\ref{s:example} we show the
results of a test case, illustrating the computation of the
preliminary orbits for a numbered asteroid whose orbit is well known,
while in Section~\ref{s:simul} we investigate the performance of the
method for a large database of simulated observations.

\section{Attributables}
\label{s:attrib}
Let $(\rho,\alpha,\delta) \in \R^+ \times [-\pi,\pi) \times
(-\pi/2,\pi/2)$ be spherical coordinates for the topocentric position
of a solar system body.  The angular coordinates $(\alpha, \delta)$ are
defined by a topocentric coordinate system that can be arbitrarily
selected.  Usually, in the applications, $\alpha$ is the right
ascension and $\delta$ the declination with respect to an equatorial
coordinate system (e.g., J2000).

Given a short arc of observations of a celestial body $(t_i, \alpha_i,
\delta_i)$, for $i=1\ldots m$ with $m\geq 2$, it is often possible to
compute an attributable,\footnote{The name refers to the possibility
of attributing the observations of the short arc to an already known
orbit.}  that is a vector
\[
\Att = (\alpha,\delta,\dot\alpha,\dot\delta) \in [-\pi,\pi) \times
(-\pi/2,\pi/2) \times \R^2\,,
\]
representing the angular position and velocity of the body at a mean
time $\bar t$ in the selected coordinates (see \cite{ident4}).
Usually we choose $\bar t$ as the mean $(\sum_i t_i)/m$.  The
attributable is computed by a polynomial fit, typically linear or
quadratic, and the observations used in the computation need to be
made by the same observatory. If the observations are enough, i.e.
$m\geq 2$ for a linear fit, $m\geq 3$ for a quadratic one, then we can
compute also a covariance matrix $\Gamma_{\Att}$, representing the
uncertainty of the attributable.  Note that the topocentric distances
$\rho_i$ at times $t_i$ are completely unknown.

\noindent We introduce the heliocentric position and velocity of the body at
time $\bar{t}$
\begin{equation}
\br = \q + \rho\hrho\,,\hskip 1cm \dbr = \dq + \dot{\rho}\hrho +
\rho(\hrhoa \alphadot + \hrhod \deltadot)\,,
\label{r_rdot}
\end{equation}
with $\rho$, $\rhodot$ the
topocentric distance and the radial velocity, and with $\hrho$,
$\hrhoa$, $\hrhod$ the observation direction and its partial
derivatives with respect to $\alpha$ and $\delta$.

The vectors $\q, \dq$ represent the heliocentric position and velocity
of the observer on the Earth. The observer position is know as a
function of time, but for consistency, if the attributable is computed
by a fit to polynomials with low degree, the values $\q(\bar t),
\dq(\bar t)$ need to be computed by the same interpolation.  Therefore
we make a quadratic fit with the actual geocentric positions
$\q(t_i)-\q_\oplus(t_i)$ at the times of the individual observations
($\q_\oplus$ is the heliocentric position of the Earth center) to
obtain the interpolating function $\q_{obs}(t)$; then we take $\q(\bar
t) = \q_\oplus(\bar t) + \q_{obs}(\bar t)$ and $\dq(\bar t) =
\dot\q_\oplus(\bar t) + \dot\q_{obs}(\bar t)$.
This method was suggested by Poincar\'e in
\cite{poincare}, and it is important to obtain preliminary orbits of
better quality, see \cite{miletal08}.

In rectangular coordinates we have
\[
\begin{array}{l}
\hrho = (\cos\alpha\cos\delta, \sin\alpha\cos\delta, \sin\delta)\,, \cr
\hrhoa = (-\sin\alpha\cos\delta, \cos\alpha\cos\delta, 0)\,, \cr
\hrhod = (-\cos\alpha\sin\delta, -\sin\alpha\sin\delta, \cos\delta)\ . \cr
\end{array}
\]
These vectors form an orthogonal system, in particular
\[
\vert\hrho\vert = \vert\hrhod\vert = 1\,,\qquad
\vert\hrhoa\vert = \cos\delta\,,\qquad
\hrho\cdot\hrhoa = \hrho\cdot\hrhod = \hrhoa\cdot\hrhod = 0\,,
\]
where the dot indicates the Euclidean scalar product and $|\cdot|$ the
corresponding norm.

For later reference we introduce the orthonormal basis $\{\hrho, \hv, \hn\}$
adapted to the apparent path 
$\hrho=\hrho(t)$ 
of the observed body on the celestial sphere: we define the unit vector $\hv$
by the relation
\[
\frac{d}{dt}\hat{\brho} = \eta\;\hv\,,
\]
where $\eta = \sqrt{\dot\alpha^2\cos^2\delta +\dot\delta^2}$ and is
called {\em proper motion}.  Moreover we set $\hn=\hat{\brho}\times
\hv$.

\section{Linkage by the two-body integrals}
\label{s:linkage}

Given two attributables $\Att_1, \Att_2$ at different epochs
$\bar{t}_1, \bar{t}_2$, in the hypothesis that they belong to the same
observed body, we write down polynomial equations for the topocentric
distance and radial velocity of the body at the two epochs by using the
angular momentum and the energy integrals.
%
%

\subsection{Angular momentum and Energy}
\label{ss:AM_energy}

For a given attributable $\Att$ the angular momentum vector (per unit mass)
can be written as a polynomial function of the radial distance and velocity
$\rho,\rhodot$:
\[
\cc(\rho,\rhodot) = \rr \times \dot{\rr} = \DD \rhodot + \EE \rho^2 + \FF \rho
+ \GG\,,
\]
where
\[
\begin{array}{l}
\DD = \q\times\hrho\,,\cr
\EE = \alphadot\hrho\times\hrhoa + \deltadot\hrho\times\hrhod = \eta\hn\,,\cr
\FF = \alphadot\q\times\hrhoa + \deltadot\q\times\hrhod + \hrho\times\dq\,,\cr
\GG = \q\times\dq\,,\cr
\end{array}
\]
depend only on the attributable $\Att$ and on the motion of the
observer $\q, \dq$ at the time $\bar t$ of the attributable.  For the
given $\Att$ we can also write the two-body energy as a function of
$\rho,\rhodot$, as in \cite{miletal04}
\[
2{\cal E}(\rho,\rhodot) = \rhodot^2 + c_1\rhodot + c_2\rho^2 + c_3\rho + c_4 -
\frac{2k^2}{\sqrt{\rho^2+c_5\rho+c_0}}\,,
\]
where $k$ is Gauss' constant and
\[
\begin{array}{lll}
c_0 = |\q|^2\,,
&\qquad c_1 = 2\; \dot\q\cdot\hrho\,,
&\qquad c_2 = \eta^2\,,\cr
c_3 = 2(\alphadot\; \dot\q\cdot\hrhoa + \deltadot\; \dot\q\cdot\hrhod)\,,
&\qquad c_4 = |\dot\q|^2\,,
&\qquad c_5 = 2\; \q\cdot\hrho\,,\cr 
\end{array}
\]
depend only on $\Att, \q, \dq$.

\subsection{Equating the integrals}

Now we take two attributables $\Att_1 =
(\alpha_1,\delta_1,\alphadot_1,\deltadot_1)$, $\Att_2 =
(\alpha_2,\delta_2,\alphadot_2,\deltadot_2)$ at epochs $\bar{t}_1,
\bar{t}_2$; we shall use the notation of Section~\ref{ss:AM_energy},
with index $1$ or $2$ referring to the epoch.  If $\Att_1$, $\Att_2$
correspond to the same physical object, then the angular momentum
vectors at the two epochs must coincide:
\begin{equation}
\DD_1\rhodot_1 - \DD_2\rhodot_2 = \JJ(\rho_1,\rho_2)\,,
\label{eq_AM}
\end{equation}
where 
\[
\JJ(\rho_1,\rho_2) = \EE_2\rho_2^2 - \EE_1\rho_1^2 + \FF_2\rho_2 -
\FF_1\rho_1 + \GG_2 - \GG_1\ .
\]
Relation (\ref{eq_AM}) is a system of three equations in the four unknowns
$\rho_1,\dot{\rho}_1,\rho_2,\dot{\rho}_2$, with constraints
\[
\rho_1>0\,,\ \rho_2>0\ .
\]
By scalar multiplication of (\ref{eq_AM}) with $\DD_1\times\DD_2$ we eliminate
 the variables $\rhodot_1, \rhodot_2$ and obtain the equation
\begin{equation}
\DD_1\times\DD_2\cdot\JJ(\rho_1,\rho_2) = 0\ .
\label{eq_AM_scal}
\end{equation}

\noindent The left hand side in (\ref{eq_AM_scal}) is a quadratic form
in the variables $\rho_1,\rho_2$; we write it as
\begin{equation}
q(\rho_1,\rho_2) \stackrel{\rm def}{=} q_{20}\rho_1^2 + q_{10}\rho_1 +
q_{02}\rho_2^2 + q_{01}\rho_2 + q_{00}\,,
\label{quad_form}
\end{equation}
with
\[
\begin{array}{l}
q_{20} = -\EE_1\cdot\DD_1\times\DD_2\,,\cr
q_{10} = -\FF_1\cdot\DD_1\times\DD_2\,,\cr
\end{array}
\hskip 1cm
\begin{array}{l}
q_{02} = \EE_2\cdot\DD_1\times\DD_2\,,\cr
q_{01} = \FF_2\cdot\DD_1\times\DD_2\,,\cr
\end{array}
\]
\[
q_{00} = (\GG_2-\GG_1)\cdot\DD_1\times\DD_2\ .
\]
Equation (\ref{quad_form}) defines a conic section in the
$(\rho_1, \rho_2)$ plane, with symmetry axes parallel to the
coordinate axes. Since the directions of $\EE_1, \EE_2$ correspond to
$\hn_1, \hn_2$, for $|\bar t_2-\bar t_1|$
small enough the angle between these two directions is small and the
coefficients $q_{20}, q_{02}$ have opposite signs, thus in this case
(\ref{quad_form}) defines a hyperbola.

\smallbreak
We can compute the radial velocities $\rhodot_1, \rhodot_2$ by
vector multiplication of (\ref{eq_AM}) with $\DD_1$ and $\DD_2$,
projecting on the direction of $\DD_1\times\DD_2$:
\begin{equation}
\rhodot_1(\rho_1,\rho_2) =
\frac{(\JJ\times\DD_2)\cdot(\DD_1\times\DD_2)}
{|\DD_1\times\DD_2|^2}\,,
\hskip 0.4cm
\rhodot_2(\rho_1,\rho_2) =
\frac{(\JJ\times\DD_1)\cdot(\DD_1\times\DD_2)}
{|\DD_1\times\DD_2|^2}\ .
\label{rhojdot}
\end{equation}

For the given $\Att_1, \Att_2$ we can also equate the corresponding
two-body energies ${\cal E}_1, {\cal E}_2$. We use the expressions of
$\rhodot_1(\rho_1, \rho_2), \rhodot_2(\rho_1, \rho_2)$ above and
substitute them into ${\cal E}_1={\cal E}_2$, thus we obtain
\begin{equation}
{\cal F}_1(\rho_1,\rho_2) - \frac{2k^2}{\sqrt{{\cal G}_1(\rho_1)}} = 
{\cal F}_2(\rho_1,\rho_2) - \frac{2k^2}{\sqrt{{\cal G}_2(\rho_2)}}\,,
\label{energyeq_nosquare}
\end{equation}
for some polynomial functions ${\cal F}_1(\rho_1,\rho_2)$, ${\cal
F}_2(\rho_1,\rho_2)$, ${\cal G}_1(\rho_1)$, ${\cal G}_2(\rho_2)$ with
degrees $\deg({\cal F}_1) = \deg({\cal F}_2) = 4$ and $\deg({\cal
G}_1) = \deg({\cal G}_2) = 2$.  By squaring we have
\begin{equation}
({\cal F}_1-{\cal F}_2)^2{\cal G}_1{\cal G}_2 - 4k^4({\cal G}_1+{\cal
G}_2) = - 8k^4\sqrt{{\cal G}_1{\cal G}_2}\ .
\label{energyeq_fstsquare}
\end{equation}
Squaring again we obtain the polynomial equation
\begin{equation}
p(\rho_1,\rho_2) \stackrel{\rm def}{=} \left[({\cal F}_1-{\cal F}_2)^2{\cal
G}_1{\cal G}_2 - 4k^4({\cal G}_1+{\cal G}_2)\right]^2 - 64k^8{\cal G}_1{\cal
G}_2 = 0 \,,
\label{eq_energies}
\end{equation}
with total degree 24. Some spurious solutions may have been added as a
result of squaring expressions with unknown sign.

Note that, if the observations were made from the center of the Earth,
$\GG_i\ (i=1,2)$ would be the angular momentum of the Earth at epochs
$\bar{t}_1$, $\bar{t}_2$, thus $\GG_1=\GG_2$ and $q_{00} = 0$.  With
this simplifying assumption $\rho_1=\rho_2=0$ is a solution of the
system $q(\rho_1,\rho_2) = p(\rho_1,\rho_2) = 0$ that corresponds to
the Earth center and therefore is not acceptable.  This solution
also appears in the geocentric version of the method of Laplace for a
preliminary orbit from 3 observations.  Actually we use topocentric
observations, for which the zero solution is replaced by one with both
$\rho_1$ and $\rho_2$ very small.

\subsection{Degenerate cases}
The quadratic form (\ref{quad_form}) degenerates into a linear function when
\[
\EE_1\cdot\DD_1\times\DD_2 = \EE_2\cdot\DD_1\times\DD_2 = 0\ .
\]
A simple computation shows that
\begin{eqnarray*}
&&\EE_1\cdot\DD_1\times\DD_2 = \eta_1
(\hn_1\cdot\q_1)(\hrho_1\times\hrho_2\cdot\q_2)\,,\\
&&\EE_2\cdot\DD_1\times\DD_2 = \eta_2
(\hn_2\cdot\q_2)(\hrho_1\times\hrho_2\cdot\q_1)\,,
\end{eqnarray*}
thus, assuming that the proper motions $\eta_1$, $\eta_2$ do not vanish and
setting $\n12 = \hrho_1\times\hrho_2$, the degeneration occurs when 
either  $\n12$ vanishes (C0) or at least
one the following relations holds:
\[
\begin{array}{ll}
\hn_1\cdot\q_1 = \hn_2\cdot\q_2 = 0\,,\hskip 2cm &\mbox{(C1)}\cr
\n12\cdot\q_1 = \n12\cdot\q_2 = 0\,,\hskip 2cm   &\mbox{(C2)}\cr
\hn_1\cdot\q_1 = \n12\cdot\q_1 = 0\,,\hskip 2cm  &\mbox{(C3)}\cr
\hn_2\cdot\q_2 = \n12\cdot\q_2 = 0\ .\hskip 2cm  &\mbox{(C4)}\cr
\end{array}
\]
The interpretation of these conditions is the following: (C0) means
that $\hrho_1, \hrho_2$ point to either exactly the same or exactly
the opposite direction in the sky; (C1) means that both sets
$\{\hrho_1, \hv_1, \q_1\}$ and $\{\hrho_2, \hv_2, \q_2\}$ are
constituted of coplanar vectors; (C2) says that $\hrho_1, \hrho_2,
\q_1, \q_2$ are coplanar.  Let us discuss condition (C3):
$\hn_1\cdot\q_1=0$ means that $\q_1, \hrho_1, \hv_1$ are coplanar and
$\n12\cdot\q_1 = 0$ means that $\hrho_1, \hrho_2, \q_1$ are coplanar
as well.  If $\DD_1\neq 0 $ we obtain that the four vectors $\q_1,
\hrho_1, \hv_1, \hrho_2$ all lie in the same plane.  In particular
(C3) implies that $\hrho_2$ belongs to the great circle defined by the
intersection of the plane generated by $\hrho_1, \hv_1$ with the
celestial sphere.  This degeneration condition can be compared with
the failure condition of the classical orbit determination methods
with three observations by Gauss and Laplace~\cite{plummer}, due to
vanishing of the curvature in the apparent path of the observed body
on the celestial sphere.  The discussion of condition (C4) is similar
to the previous one and corresponds to the coplanarity of $\q_2,
\hrho_2, \hv_2, \hrho_1$.

\section{Computation of the solutions}
\label{s:algor}

In this section we introduce two different methods to search for the
solutions of the semi-algebraic problem
\begin{equation}
\left\{
\begin{array}{l}
p(\rho_1,\rho_2) = 0\cr
q(\rho_1,\rho_2) = 0\cr
\end{array}
\right.\,,
\hskip 1cm
\rho_1, \rho_2>0
\label{intersec}
\end{equation}
for the polynomials $p, q$ introduced in (\ref{eq_energies}),
(\ref{quad_form}) respectively.  Moreover we explain the full
procedure for the computation of the preliminary orbits and
introduce compatibility conditions to decide whether the attributables
used to define the problem are related to the same solar system body.

\subsection{Computation of the resultant via DFT}
\label{ss:DFT}

The first method consists in writing the resultant (see \cite{cox}) of
$p$ and $q$ with respect to one variable, say $\rho_1$.  In this way
we find a univariate polynomial in the $\rho_2$ variable whose real
positive roots are the only possible $\rho_2$-components of a solution
of (\ref{intersec}).
By grouping the monomials with the same power of $\rho_1$ we can write
\begin{equation}
p(\rho_1,\rho_2) = \sum_{j=0}^{20} a_j(\rho_2)\;\rho_1^j\,,
\hskip 1cm\mbox{where}
\label{enerpoly}
\end{equation}
\[
\deg(a_j)=\left\{
\begin{array}{lll}
20 &\mbox{ for }j=0\ldots 4  &\cr
24-(j+1) &\mbox{ for }j=2k-1 &\mbox{ with } k\ge 3\cr
24-j &\mbox{ for }j=2k       &\mbox{ with } k\ge 3\cr
\end{array}
\right.
\]
and
\begin{equation}
q(\rho_1,\rho_2) = b_2\;\rho_1^2 + b_1\;\rho_1 + b_0(\rho_2)
\label{AMpoly}
\end{equation}
for some univariate polynomial coefficients $a_i, b_j$ depending on $\rho_2$
(actually $b_1, b_2$ are constant).
We consider the resultant $Res(\rho_2)$ of $p,q$ with respect to
$\rho_1$: it is generically a degree 48 polynomial
defined as the determinant of the Sylvester matrix
\begin{equation}
{\tt S}(\rho_2) = 
\left(
\begin{array}{ccccccc}
a_{20} &0      &b_2    &0      &\ldots &\ldots &0      \cr
a_{19} &a_{20} &b_1    &b_2    &0      &\ldots &0      \cr
\vdots &\vdots &b_0    &b_1    &b_2    &\ldots &\vdots \cr
\vdots &\vdots &0      &b_0    &b_1    &\ldots &\vdots \cr
a_0    &a_1    &\vdots &\vdots &\vdots &b_0    &b_1    \cr
0      &a_0    &0      &0      &0      &0      &b_0    \cr
\end{array}
\right)\ .
\end{equation}
The positive real roots of $Res(\rho_2)$ are the only possible values of
$\rho_2$ for a solution $(\rho_1,\rho_2)$ of (\ref{intersec}).
We could use the resultant method to eliminate the variable $\rho_2$
by a different grouping of the terms of $p,q$:
\[
p(\rho_1,\rho_2) = \sum_{j=0}^{20} a'_j(\rho_1)\;\rho_2^j\ ,
\hskip 1cm
q(\rho_1,\rho_2) = b'_2\;\rho_2^2 + b'_1\;\rho_2 + b'_0(\rho_1)\,,
\]
where the degrees of $a'_j$ are described by the same rules as for $a_j$.


Apart from non-real and non-positive solutions, we shall see that there are
additional different reasons to discard some pairs of solutions of
(\ref{intersec}), thus we expect that the number of acceptable ones is not
large.
We use a scheme similar to \cite{gronchi02} to
compute the coefficients of the resultant $Res(\rho_2)$:
\begin{itemize}
\item[1)] evaluate $a_i(\rho_2), b_j(\rho_2)$ at the 64-th roots of
unit $\omega_k = e^{2\pi i{k\over 64}}\ , k=0\ldots 63$, by a DFT
(Discrete Fourier Transform) algorithm;

\item[2)] compute the determinant of the 64 Sylvester matrices;
by relation 
\[\det\left({\tt S}(\rho_2)|_{\rho_2=\omega_k}\right) =
\left(\det{\tt S}(\rho_2)\right)|_{\rho_2=\omega_k}
\]  
we have the values of $Res(\rho_2)$ at the 64-th roots of unit;

\item[3)] apply an IDFT (Inverse Discrete Fourier Transform)
algorithm to obtain the coefficients of $Res(\rho_2)$ from its
evaluations.

\end{itemize}

The use of the DFT and IDFT allows us to interpolate the resultant
$Res(\rho_2)$ in an efficient way. The use of numerical evaluations,
e.g. at the roots of unit, avoids the difficulty of writing a very
long symbolic expression for the resultant, that could be cumbersome
to be managed by a programming language compiler.

The complete set of complex roots of $Res(\rho_2)$, with an error
bound for each of them, are computed using the algorithm described in
\cite{bini}, which is based on simultaneous iterations. Let
$\rho_2(k)$, $k=1\ldots n\leq 48$, be the subset of the real and
positive roots of $Res(\rho_2)$.  Then for each $k$ we perform the
sequence of operations below:
\begin{itemize}
\item[4)] solve the equation $q(\rho_1,\rho_2(k)) = 0$ and compute the
  two possible values $\rho_1(k,1)$, $\rho_1(k,2)$ for $\rho_1$,
  discarding negative solutions. Then define $\rho_1(k)$ equal to
  either $\rho_1(k,1)$ or $\rho_1(k,2)$, selecting the one that gives
  the smaller value of $|p(\rho_1, \rho_2(k))|$;

\item[5)] discard spurious solutions, resulting from the squaring used
  to reduce the energy equality to the polynomial equation
  (\ref{eq_energies}). The spurious solutions are the solutions of
  (\ref{intersec}) that do not satisfy either
  (\ref{energyeq_fstsquare}) or (\ref{energyeq_nosquare});

\item[6)] compute the corresponding values of $\dot{\rho}_1(k),
  \dot{\rho}_2(k)$ by (\ref{rhojdot}) and obtain a pair of orbits
  defined by the sets $(\alpha_i, \delta_i, \alphadot_i, \deltadot_i,
  \rho_i, \rhodot_i)$ of {\em attributable elements},\footnote{The
    attributable elements are the same as spherical polar coordinates
    with their time derivatives: the coordinates are just reordered in
    such a way that the first four elements form the attributable,
    hence the name.} for $i=1,2$;

\item[7)] change from attributable elements to Cartesian heliocentric
  coordinates by relation $\br_i=\rho_i(k) \hrho_i + \q_i$ for
  $i=1,2$, and by the corresponding formula for $\dbr_i$. Note that
  the observer position $\q_i$ is not the actual $\q(t_i)$, but is
  obtained by interpolation as proposed by Poincar\'e (see
  Section~\ref{s:attrib}). Then a standard coordinate change allows us
  to obtain the related pairs of orbital elements: we shall use
  Keplerian elements $(a,e,I, \Omega, \omega, \ell)$, where $\ell$ is
  the mean anomaly\footnote{Any other set of orbital elements in which
    the first four are defined by the two-body energy and angular
    momentum can be used, e.g. cometary elements $(p_d, e, I, \Omega,
    \omega, t_p)$ where $p_d$ is the perihelion distance and $t_p$ the
    time of perihelion passage: this set would allow to handle also
    parabolic and hyperbolic orbits.}. The epochs of the orbits are
  $\tilde t_1(k), \tilde t_2(k)$, corrected by aberration due to the
  finite velocity of the light $c$: $\tilde t_i(k)= \bar{t}_i -
  \rho_i(k)/c$ for $i=1,2$.

\end{itemize}

We have implemented this algorithm in FORTRAN 90 using {\em quadruple
precision} for part of these computations, in particular the ones
related to DFT and IDFT. This feature appeared necessary to obtain
reliable results starting from our first numerical experiments.

\subsection{Normal form of the problem}
\label{ss:normal_form}

Another method to compute the solutions of (\ref{intersec}) is based
on a coordinate change to variables $(\xi_1, \xi_2)$, that allows to
perform easily the elimination of either $\xi_1$ or $\xi_2$. 
Let us set
\[
p(\rho_1,\rho_2) = \sum_{i,j=0}^{20} p_{i,j}\rho_1^i \rho_2^j\ .
\]
First we consider the affine transformation to intermediate variables
$(\zeta_1, \zeta_2)$
\[
{\cal T} : \left(
\begin{array}{c}
\rho_1\cr
\rho_2\cr
\end{array}
\right) \rightarrow
\left(
\begin{array}{c}
\zeta_1\cr
\zeta_2\cr
\end{array}
\right) = 
\left(
\begin{array}{c}
\sigma_1^{-1}\rho_1 - \tau_1\cr
\sigma_2^{-1}\rho_2 - \tau_2\cr
\end{array}
\right)\,,
\]
where, to eliminate the linear terms in (\ref{quad_form}), we set
\[
\sigma_1 \tau_1 = -\frac{q_{1,0}}{2 q_{2,0}} \stackrel{\rm def}{=} \alpha \,,
\hskip 1cm
\sigma_2 \tau_2 = -\frac{q_{0,1}}{2 q_{0,2}} \stackrel{\rm def}{=} \beta \,,
\]
so that
\[
q\circ {\cal T}^{-1}(\zeta_1,\zeta_2) = q_{2,0}\sigma_1^2\Bigl[\zeta_1^2 +
\frac{q_{0,2}\sigma_2^2}{q_{2,0}\sigma_1^2}\zeta_2^2 +
\frac{\kappa}{q_{2,0} \sigma_1^2}\Bigr]\,,\hskip 0.6cm
\mbox{with }\ \kappa = q_{0,0} - \frac{q_{1,0}^2}{4q_{2,0}} -
\frac{q_{0,1}^2}{4q_{0,2}}\ .
\]
If we set, for an arbitrary $\sigma_2\in\R$,
\[
\sigma_1 = \gamma \sigma_2\,,
\hskip 1cm
\gamma = \sqrt{-\frac{q_{0,2}}{q_{2,0}}}
\]
we obtain
\[
q\circ {\cal T}^{-1} (\zeta_1,\zeta_2) =
q_{2,0}\sigma_1^2 \left[\zeta_1^2 -\zeta_2^2 -2\cstar\right]\,,
\hskip 1cm
\mbox{with }\ \cstar = -\frac{\kappa}{2\; q_{2,0} \sigma_1^2}\ .
\]
We already observed that, for $|\bar t_2-\bar t_1|$ small enough,
$q_{02}$ and $q_{20}$ have opposite signs, hence in this case the
variable change ${\cal T}$ is real.  However, in general, we have to
consider ${\cal T}$ as a transformation of the complex domain $\C^2$.
We also have
\[
p\circ {\cal T}^{-1}(\zeta_1,\zeta_2) = \sum_{i,j=0}^{20}
\tilde{p}_{i,j} \zeta_1^i \zeta_2^j\,,
\]
where
\begin{equation}
\tilde{p}_{i,j} = \sigma_2^{i+j} \gamma^i \sum_{h=i}^{20}
\sum_{k=j}^{20} p_{h,k} 
\left(\begin{array}{c} h\cr i\cr \end{array}\right)
\left(\begin{array}{c} k\cr j\cr \end{array}\right)
\alpha^{h-i} \beta^{k-j}\,,
\label{ptilde}
\end{equation}
and $\alpha,\beta,\gamma$ depend only on the coefficients of
$q(\rho_1,\rho_2)$.

\noindent Now we apply a rotation of angle $\pi/4$ to pass to the
$(\xi_1, \xi_2)$ variables:
\[
{\cal R}: \left(
\begin{array}{c}
\zeta_1\cr
\zeta_2\cr
\end{array}
\right) \rightarrow 
\left(
\begin{array}{c}
\xi_1\cr
\xi_2\cr
\end{array}
\right) = 
\left[
\begin{array}{cc}
\cos(\frac{\pi}{4})   &-\sin(\frac{\pi}{4})\cr
\sin(\frac{\pi}{4})   &\cos(\frac{\pi}{4})\cr
\end{array}
\right]
\left(
\begin{array}{c}
\zeta_1\cr
\zeta_2\cr
\end{array}
\right)\ .
\]
We have
\begin{eqnarray*}
q\circ{\cal T}^{-1}\circ{\cal R}^{-1} (\xi_1,\xi_2) &=&
2\;q_{2,0}\sigma_1^2\left[\xi_1 \xi_2 - \cstar\right]\,,\\
p\circ{\cal T}^{-1}\circ{\cal
R}^{-1} (\xi_1,\xi_2) &=& \sum_{i,j=0}^{24} p^\star_{i,j} \xi_1^i
\xi_2^j\,,
\end{eqnarray*}
where
\begin{equation}
p^\star_{i,j} = 
\left\{
\begin{array}{ll}
\bar{p}^\star_{i+j,i}\qquad &\mbox{ if } i+j\leq 24\cr
0 &\mbox{ if } i+j > 24\cr
\end{array}
\right.
\end{equation}
and
\begin{eqnarray*}
\bar{p}^\star_{m,n} &=& 
\sum_{h+k=m} \tilde{p}_{h,k} 
\sum_{i+j=n}
\left(\begin{array}{c} h\cr i\cr \end{array}\right)
\left(\begin{array}{c} k\cr j\cr \end{array}\right)
\frac{(-1)^j}{2^{(h+k)/2}} = 
%
\sum_{h=0}^m \tilde{p}_{h,m-h} 
\sum_{i=0}^n
\left(\begin{array}{c} h\cr i\cr \end{array}\right)
\left(\begin{array}{c} m-h\cr n-i\cr \end{array}\right)
\frac{(-1)^{n-i}}{2^{m/2}} = \\    
%
&=& \sum_{h=\max\{m-20,0\}}^{\min\{m,20\}} \tilde{p}_{h,m-h} 
\sum_{i=0}^n
\left(\begin{array}{c} h\cr i\cr \end{array}\right)
\left(\begin{array}{c} m-h\cr n-i\cr \end{array}\right)
\frac{(-1)^{n-i}}{2^{m/2}} \ .
\end{eqnarray*}
The last equality is obtained taking into account that
$\tilde{p}_{h,m-h} = 0\ \ \mbox{ for }\ \ h>20\ \ \mbox{ or }\ \ m-h>20$.
%
Using the relation $\xi_1\xi_2=\cstar$ we can consider in place of
$p\circ{\cal T}^{-1}\circ{\cal R}^{-1} (\xi_1,\xi_2)$
the polynomial
\begin{eqnarray*}
p^\star (\xi_1,\xi_2) &=& 
\sum_{\tiny
\begin{array}{c}
{h,k=0}\cr
{h>k}\cr
\end{array}}^{24} \phkstar \cstar^k \xi_1^{h-k} +
\sum_{h=0}^{24} \phhstar \cstar^h +
\sum_{\tiny
\begin{array}{c}
{h,k=0}\cr
{h<k}\cr
\end{array}}^{24} \phkstar \cstar^h \xi_2^{k-h} = \\
&=& \sum_{j=1}^{24}\Biggl(
\sum_{\tiny
\begin{array}{c}
{h,k=0}\cr
{h-k=j}\cr
\end{array}}^{24} \phkstar \cstar^{h-j}\Biggr) \xi_1^j +
\sum_{j=1}^{24}\Biggl(
\sum_{\tiny
\begin{array}{c}
{h,k=0}\cr
{k-h=j}\cr
\end{array}}^{24} \phkstar \cstar^{k-j}\Biggr) \xi_2^j +
\sum_{h=0}^{24} \phhstar \cstar^h  = \\
&=& A_{24}\,\xi_1^{24} + \ldots + A_1\xi_1 +
A_0\,(\xi_2) \,,
\end{eqnarray*}
with 
\[
A_j = \sum_{\tiny
\begin{array}{c}
{h,k=0}\cr
{h-k=j}\cr
\end{array}}^{24} \phkstar \cstar^{h-j} = 
\sum_{h=j}^{24}
p_{h,h-j}^\star \cstar^{h-j}\,,
\hskip 0.5cm 
j=1\ldots 24\,,
\]
\[
A_0(\xi_2) = B_{24}\,\xi_2^{24} + \ldots +B_1\,\xi_2 + B_0\,, 
\]
\[
%
B_j = \sum_{\tiny
\begin{array}{c}
{h,k=0}\cr
{k-h=j}\cr
\end{array}}^{24} \phkstar \cstar^{k-j} = 
\sum_{k=j}^{24}
p_{k-j,k}^\star \cstar^{k-j} \,, 
\hskip 0.5cm j=0\ldots 24\ .
\]

\noindent 
We consider the algebraic problem in normal form
\begin{equation}
\left\{
\begin{array}{l}
p^\star(\xi_1,\xi_2) = 0 \cr 
\xi_1\xi_2 - \cstar = 0\cr
\end{array}
\right.\ .
\label{normal_form}
\end{equation}
%
%
In this case we have to consider all the solutions of
(\ref{normal_form}), not only the ones with real and positive
components.

\noindent If $\cstar=0$, then the solutions $(\xi_1,\xi_2)$ of
(\ref{normal_form}) are of the form $(\xi_1(k),0)$ or $(0,\xi_2(k))$,
where $\xi_1(k), \xi_2(k)$, $k=1\ldots 24$ are the roots of
$A_{24}\,\xi_1^{24} + \ldots + A_1\xi_1 + B_0$ and $B_{24}\,\xi_2^{24}
+ \ldots + B_1\xi_2 + B_0$ respectively.  

\noindent If $\cstar\neq 0$, using the
relation $\xi_1\xi_2=\cstar$ we can eliminate one variable, say
$\xi_1$, from $p^\star$.  Thus we obtain the univariate polynomial
\[
\mathfrak{p}(\xi_2) = \sum_{k=0}^{48} \mathfrak{p}_k \xi_2^k\,,
\hskip 1cm
\mbox{with}\ \
\mathfrak{p}_k = 
\left\{\begin{array}{ll}
 A_{24-k}\cstar^{24-k}\,,\hskip 0.3cm &0\leq k\leq 23\cr 
 B_{k-24}\,, &24\leq k\leq 48\cr
\end{array}\right.\ .
\]
We compute all the complex roots $\xi_2(k), k=1\ldots 48$ of
$\mathfrak{p}(\xi_2)$ by the algorithm in \cite{bini};  then for each
$k$ we define the other component of the solution by
\[
\xi_1(k) = \frac{\cstar}{\xi_2(k)} \ .
\]
Given all the complex solutions of (\ref{normal_form}) we compute the
corresponding points in the $(\rho_1,\rho_2)$ plane by
\[
\left( \rho_1(k), \rho_2(k) \right) =
{\cal T}^{-1}\circ{\cal R}^{-1} \left( \xi_1(k), \xi_2(k)
\right)\,,\hskip 1cm k=1\ldots 48\,,
\]
discarding the ones with non-real or non-positive components.
At this point the preliminary orbits can be computed following the
same steps 5), 6), 7) of the algorithm explained in
Subsection~\ref{ss:DFT}.

From a few experiments performed this method seems to require more
than quadruple precision because of the complicated formulae defining
the transformation used to obtain the normal form (\ref{normal_form}).
Thus the advantage in the simple elimination of the variable $\xi_1$
must be balanced with the introduction of heavier computations.

\subsection{Compatibility conditions}

The knowledge of the angular momentum vector and of the energy at a
given time allows us to compute the Keplerian elements
\[
a,e,I,\Omega\ .
\]
In fact the semimajor axis $a$ and the eccentricity $e$ can be computed from
the energy and the size of the angular momentum through the relations
\[
\mathcal{E} = -\frac{k^2}{2a}\,,
\hskip 1cm
\Vert\cc\Vert=k\sqrt{a(1-e^2)}\,;
\]
the longitude of the node $\Omega$ and the inclination $I$ are obtained from
the direction of the angular momentum
\[
\hat{\cc} = (\sin\Omega\sin I, -\cos\Omega\sin I, \cos I)\ .
\]
The two attributables $\Att_1, \Att_2$ at epochs $\bar{t}_1,
\bar{t}_2$ give 8 scalar data, thus the problem is over-determined.
From a non-spurious pair $(\tilde{\rho}_1,\tilde{\rho}_2)$, solution
of (\ref{intersec}), we obtain the same values of $a,e,I,\Omega$ at
both times $\tilde{t}_i, i=1,2$, but we must check that the orbit is
indeed the same, that is check the compatibility conditions
\begin{equation}
\omega_1 = \omega_2\,,
\hskip 1.2cm
\ell_1 = \ell_2 + n(\tilde{t}_1-\tilde{t}_2)\,,
\label{compatible}
\end{equation}
where $\omega_1, \omega_2$ and $\ell_1, \ell_2$ are the arguments of
perihelion and the mean anomalies of the body at times $\tilde{t}_1,
\tilde{t}_2$ and $n = k a^{-3/2}$ is the mean motion, which is the
same for the two orbits.  The first of conditions (\ref{compatible})
corresponds to the use of the fifth integral of the Kepler problem,
related to Lenz-Laplace's integral vector
\[
\L = \frac{1}{k^2} \dbr \times \cc - \frac{\rr}{\vert\rr\vert}\ .
\]
Indeed the compatibility conditions (\ref{compatible}) can not be
exactly satisfied, due to both the errors in the observations and to
the planetary perturbations. Actually the latter are important only
when the observed body undergoes a close approach to some planet in
the interval between $\tilde t_1$ and $\tilde t_2$.  Thus we may be
able to discard some solutions, for which the compatibility conditions
are largely violated. Nevertheless, we need a criterion to assess
whether smaller discrepancies from the exact conditions
(\ref{compatible}) are due to the measurement uncertainty or rather
due to the fact that the two attributables do not belong to the same
physical object. This will be introduced in the next section.

%

\section{Covariance of the solutions}
\label{s:uncert}

Given a pair of attributables $\Avec = (\Att_1,\Att_2)$ with
covariance matrices $\Gamma_{\Att_1}, \Gamma_{\Att_2}$, we call $\Rvec
= (\rho_1, \rhodot_1, \rho_2, \rhodot_2)$ one of the solutions of the
equation $\bPhi(\Rvec;\Avec) = {\bf 0}$, with
\begin{equation}
\bPhi(\Rvec; \Avec) = \left(
\begin{array}{c}
\DD_1\rhodot_1 - \DD_2\rhodot_2 - \JJ(\rho_1,\rho_2) \cr
{\cal E}_1(\rho_1,\dot{\rho}_1) - {\cal E}_2(\rho_2,\dot{\rho}_2)\cr
\end{array}
\right)\ .
\label{phi_map}
\end{equation}
We can repeat what follows for each solution of $\bPhi(\Rvec;\Avec) = {\bf 0}$.

Let $\Rvec = \Rvec(\Avec)=(\Rcal_1(\Avec),\Rcal_2(\Avec))$,
where $\Rcal_i(\Avec) = (\rho_i(\Avec), \rhodot_i(\Avec))$ for
$i=1,2$.
If both the elements $(\Att_1,\Rcal_1(\Avec))$,
$(\Att_2,\Rcal_2(\Avec))$ give negative two-body energy orbits,
then we can compute the corresponding Keplerian elements
at times
\[
\tilde t_i = \tilde t_i(\Avec) = \bar t_i -
\frac{\rho_i(\Avec)}{c}\,,\qquad i=1,2
\]
through the transformation
\[
(\alpha,\delta,\alphadot,\deltadot,\rho,\rhodot) = (\Att,\Rcal) \mapsto
{\cal E}_{Kep}(\Att,\Rcal) = (a,e,I,\Omega,\omega,\ell)\ .
\]
We have, for example, a smooth function 
\[
\omega_i = \omega_i(\Avec) = \omega(\Att_i,\Rcal_i(\Avec))\,,\qquad i=1,2
\]
and similar functional relations for $a,e,I,\Omega,\ell$.  Actually,
by construction, we have $a_1=a_2, e_1=e_2, I_1=I_2,
\Omega_1=\Omega_2$: we denote by $\mathsf{a}$ the common value of
$a_1$ and $a_2$.
%
%
We use the vector differences
\[
\bDelta =(\Delta\omega, \Delta\ell)\,, \qquad
\]
where $\Delta\omega$ is the difference of the two angles $\omega_1$
and $\omega_2$, $\Delta\ell$ is the difference of the two angles
$\ell_1$ and $\ell_2 + \mathsf{n}(\tilde t_1-\tilde t_2)$ and
$\mathsf{n} = k \mathsf{a}^{-3/2}$ is the mean motion of both orbits.
Here we compute the difference of two angles
in such a way that 
it is a smooth function near a vanishing point; for example we define
$\Delta\omega= [\omega_1-\omega_2 +\pi (\mathrm{mod}\ 2\pi)] -\pi$.
With this caution, the vector $\bDelta=\bDelta(\Avec)$ represents the
discrepancy in perihelion argument and mean anomaly of the two orbits,
comparing the anomalies at the same time $\tilde{t}_1$.
We introduce the map
\[
\bPsi : \left([-\pi,\pi) \times (-\frac{\pi}{2},\frac{\pi}{2}) \times
\R^2\right)^2 \longrightarrow [-\pi,\pi) \times (-\frac{\pi}{2},\frac{\pi}{2})
\times \R^2 \times \R^+\times \R \times S^1\times S^1
\]
\[
(\Att_1, \Att_2) = \Avec \mapsto \bPsi(\Avec) = \left(\Att_1, \Rcal_1,\bDelta
\right)\,,
\]
giving the orbit $(\Att_1,\Rcal_1(\Avec))$ in attributable elements at
time $\tilde t_1$ (the epoch of the first attributable corrected by
aberration), together with the difference $\bDelta(\Avec)$ in the
angular elements, which are not constrained by the angular momentum
and the energy integrals.
By the covariance propagation rule we have
\begin{equation}
\Gamma_{\bPsi(\Avec)} = \frac{\partial \bPsi}{\partial
\Avec }\; \Gamma_\Avec \; \left[\frac{\partial \bPsi}{\partial
\Avec }\right]^T\ ,
\label{cov_prop_rule}
\end{equation}
where
\[
 \frac{\partial
\bPsi}{\partial \Avec} = \left[
\begin{array}{cc}
I &0\cr
\displaystyle\frac{\partial \Rcal_1}{\partial \Att_1} 
&\displaystyle\frac{\partial \Rcal_1}{\partial
\Att_2} \cr
\stackrel{}{\displaystyle\frac{\partial \bDelta}{\partial \Att_1}} 
&\displaystyle\frac{\partial
\bDelta}{\partial \Att_2} \cr
\end{array}
\right]
\hskip 0.5cm\mbox{ and }
\hskip 0.5cm
\Gamma_\Avec =  \left[
\begin{array}{cc}
\Gamma_{\Att_1} &0\cr
0 &\Gamma_{\Att_2}\cr
\end{array}
\right]\,,
\]
so that the covariance of $\bPsi(\Avec)$ is given by the $8\times 8$ matrix
\[
\Gamma_{\bPsi(\Avec)}
=\left[ 
\begin{array}{ccc}
\Gamma_{\Att_1}  
&\Gamma_{\Att_1,\Rcal_1}
&\Gamma_{\Att_1,\bDelta}\cr
\Gamma_{\Rcal_1,\Att_1}
&\Gamma_{\Rcal_1}
&\Gamma_{\Rcal_1,\bDelta}\cr
\Gamma_{\bDelta,\Att_1}
&\Gamma_{\bDelta,\Rcal_1}
&\Gamma_{\bDelta}\cr
\end{array}
\right]\ ,
\]
where
\[
\Gamma_{\Att_1,\Rcal_1} = \Gamma_{\Att_1}\left[\frac{\partial
\Rcal_1}{\partial\Att_1}\right]^T\,,
\hskip 0.5cm
\Gamma_{\Att_1,\bDelta} = \Gamma_{\Att_1}\left[\frac{\partial \bDelta}{\partial\Att_1}\right]^T\,,
\]
\[
\Gamma_{\Rcal_1,\bDelta} = \frac{\partial \Rcal_1}{\partial\Att_1}
\Gamma_{\Att_1}\left[\frac{\partial \bDelta}{\partial\Att_1}\right]^T +
\frac{\partial \Rcal_1}{\partial\Att_2}
\Gamma_{\Att_2}\left[\frac{\partial \bDelta}{\partial\Att_2}\right]^T \,,
\]
\[
\Gamma_{\Rcal_1,\Att_1} = \Gamma_{\Att_1,\Rcal_1}^T\,,
\hskip 0.5cm  
\Gamma_{\bDelta,\Att_1}=\Gamma_{\Att_1,\bDelta}^T\,,
\hskip 0.5cm
\Gamma_{\bDelta,\Rcal_1}^T=\Gamma_{\Rcal_1,\bDelta}^T\,,
\]
and
\[
\Gamma_{\Att_1} = \frac{\partial \Att_1}{\partial
\Avec}\Gamma_{\Avec} \left[\frac{\partial \Att_1}{\partial
\Avec}\right]^T\,,
\hskip 0.5cm
\Gamma_{\Rcal_1} = \frac{\partial \Rcal_1}{\partial
\Avec}\Gamma_{\Avec} \left[\frac{\partial \Rcal_1}{\partial
\Avec}\right]^T\,,
\hskip 0.5cm
\Gamma_{\Delta_{1,2}} = \frac{\partial \Delta_{1,2}}{\partial
\Avec}\Gamma_{\Avec} \left[\frac{\partial \Delta_{1,2}}{\partial
\Avec}\right]^T\ .
\]
The matrices $\frac{\partial \Rcal_i}{\partial\Att_j}, i,j=1,2$, can be
computed from the relation
\[
\frac{\partial \Rvec}{\partial \Avec }(\Avec) = - \left[ \frac{\partial
\bPhi}{\partial \Rvec}(\Rvec(\Avec),\Avec) \right]^{-1}
\frac{\partial \bPhi}{\partial\Avec}(\Rvec(\Avec),\Avec)\ .
\]
We also have
\begin{eqnarray*}
&&\frac{\partial \Delta\omega}{\partial \Att_1} = 
\frac{\partial
\omega}{\partial \Att}(\Att_1,\Rcal_1(\Avec)) + \frac{\partial
\omega}{\partial \Rcal}(\Att_1,\Rcal_1(\Avec))\frac{\partial
\Rcal_1(\Avec)}{\partial\Att_1} -
\frac{\partial \omega}{\partial \Rcal}(\Att_2,\Rcal_2(\Avec))\frac{\partial
\Rcal_2(\Avec)}{\partial\Att_1}\,,\\
&&\frac{\partial \Delta\omega}{\partial \Att_2} = 
\frac{\partial \omega}{\partial \Rcal}(\Att_1,\Rcal_1(\Avec))\frac{\partial
\Rcal_1(\Avec)}{\partial\Att_2} -
\frac{\partial
\omega}{\partial \Att}(\Att_2,\Rcal_2(\Avec)) - \frac{\partial
\omega}{\partial \Rcal}(\Att_2,\Rcal_2(\Avec))\frac{\partial
\Rcal_2(\Avec)}{\partial\Att_2} 
\end{eqnarray*}
and
\begin{eqnarray*}
\frac{\partial \Delta\ell}{\partial \Att_1} &=& 
\frac{\partial \ell}{\partial \Att}(\Att_1,\Rcal_1(\Avec)) + \frac{\partial
\ell}{\partial \Rcal}(\Att_1,\Rcal_1(\Avec))\frac{\partial
\Rcal_1(\Avec)}{\partial\Att_1} -
\frac{\partial \ell}{\partial \Rcal}(\Att_2,\Rcal_2(\Avec))\frac{\partial
\Rcal_2(\Avec)}{\partial\Att_1} +\\
&+& \frac{3}{2}\frac{\mathsf{n}}{\mathsf{a}} 
\biggl[\frac{\partial a}{\partial \Att}(\Att_1,\Rcal_1(\Avec)) +
\frac{\partial a}{\partial \Rcal}(\Att_1,\Rcal_1(\Avec))\frac{\partial
\Rcal_1(\Avec)}{\partial\Att_1} \biggr]
[\tilde t_1(\Avec)-\tilde t_2(\Avec)] + \\ 
&+&\frac{\mathsf{n}}{c}\left[\frac{\partial\rho_1}{\partial \Att_1}(\Avec) -
\frac{\partial\rho_2}{\partial \Att_1}(\Avec)\right]\,,\\
&&\\
\frac{\partial \Delta\ell}{\partial \Att_2} &=& 
\frac{\partial
\ell}{\partial \Rcal}(\Att_1,\Rcal_1(\Avec))\frac{\partial
\Rcal_1(\Avec)}{\partial\Att_2} -
\frac{\partial \ell}{\partial \Att}(\Att_2,\Rcal_2(\Avec)) 
- \frac{\partial \ell}{\partial \Rcal}(\Att_2,\Rcal_2(\Avec))\frac{\partial
\Rcal_2(\Avec)}{\partial\Att_2} +\\
&+& \frac{3}{2}\frac{\mathsf{n}}{\mathsf{a}} 
\biggl[\frac{\partial a}{\partial
\Att}(\Att_2,\Rcal_2(\Avec)) + 
\frac{\partial a}{\partial \Rcal}(\Att_2,\Rcal_2(\Avec))\frac{\partial
\Rcal_2(\Avec)}{\partial\Att_2}\biggr]
[\tilde t_1(\Avec)- \tilde t_2(\Avec)] +\\ 
&+& \frac{\mathsf{n}}{c}\left[\frac{\partial\rho_1}{\partial \Att_2}(\Avec) -
\frac{\partial\rho_2}{\partial \Att_2}(\Avec)\right]\ .
\end{eqnarray*}

\subsection{Identification of attributables}

One important step is to decide if trying to link $\Att_1$, $\Att_2$
has produced at least one reliable orbit, so that we can state
the two sets of observations defining the $\Att_i, i=1,2$ may belong
to one and the same solar system body.
Neglecting the unavoidable errors in the observations and the
approximations made both with the interpolation to compute $\Att_1$,
$\Att_2$ and with the use of a two-body model, if the observations
belong to the same solar system body, then $\Delta_{1,2}(\Avec) = {\bf
0}$. We need to check whether the failure of this condition is within the
acceptable range of values which is statistically expected to be
generated by the errors in the available observations.

\noindent The marginal covariance matrix of the compatibility
conditions is
\[
\Gamma_{\Delta_{1,2}} = \frac{\partial \Delta_{1,2}}{\partial
\Avec}\Gamma_{\Avec} \left[\frac{\partial \Delta_{1,2}}{\partial
\Avec}\right]^T\ .
\]
The inverse matrix $C^{\Delta_{1,2}} = \Gamma^{-1}_{\Delta_{1,2}}$ defines a
norm $\Vert\cdot\Vert_{\star}$ in the $(\Delta\omega, \Delta\ell)$ plane,
allowing to test an identification between the attributables $\Att_1, \Att_2$:
the test is
\begin{equation}
\Vert \Delta_{1,2}\Vert_{\star}^2 = \Delta_{1,2} C^{\Delta_{1,2}}
\Delta_{1,2}^T\leq \chi_{max}^2\ ,
\label{control}
\end{equation}
where $\chi_{max}$ is a control parameter. The value of the control
could be selected on the basis of $\chi^2$ tables, if we could assume
that the observations errors are Gaussian and their standard
deviations, mean values and correlations were known. Since this
hypothesis is not satisfied in practice, the control value
$\chi_{max}$ needs to be selected on the basis of large scale tests.
Note that, for each pair of attributables, more than one preliminary
orbit computed with the method of Section~\ref{s:algor} could pass the
control (\ref{control}); thus we can have alternative preliminary
orbits.

\subsection{Uncertainty of the orbits}

The methods explained in Section~\ref{s:algor} also allow to assign an
uncertainty to the preliminary orbits that we compute from the two
attributables. A solution $(\Att_1,\Rcal_1(\Avec))$, in attributable
elements, has the marginal covariance matrix
\[
\left[
\begin{array}{cc}
\Gamma_{\Att_1}  &\Gamma_{\Att_1,\Rcal_1}\cr
\Gamma_{\Rcal_1, \Att_1} &\Gamma_{\Rcal_1}\cr
\end{array}
\right]\ .
\]
The preliminary orbits obtained by the other available algorithms do
not produce a nondegenerate covariance matrix: this is usually
computed in the differential correction step of the orbit
determination procedure.  With the algorithm of \cite{miletal05} a
covariance matrix may be defined, but it is not positive definite.
The advantage of having a covariance matrix already from the
preliminary orbit step could be important in two ways.  First, the
covariance matrix describes a confidence ellipsoid where a two-body
orbit, compatible with the observations and their errors, can be
found.  The size of this ellipsoid can provide useful hints on the
difficulty of the differential corrections procedure.
Second, even if the differential corrections are divergent, the
covariance matrix of the preliminary orbit can be used to compute a
prediction with confidence region, allowing for a planned recovery,
for assessment of impact risk, and so on.

\section{A test case}
\label{s:example}

We show a test of the linkage procedure using the attributables
\begin{eqnarray*}
\Att_1 &=& (0.2872656, 0.1106342, -0.00375115, -0.00167695)\,,\\
\Att_2 &=& (0.2820817, 0.1086542,  0.00514465,  0.00215975) 
\end{eqnarray*}
of the asteroid ($101878$) $1999$ NR$_{23}$ at epochs $\bar{t}_1 =
54000$, $\bar{t}_2 = 54109$ respectively (time in MJD). The values of
the components of $\Att_1, \Att_2$ are in radians/radians per
day. They have been computed from two groups of observations,
separated by more than 100 days, made from two different
observatories: Mauna Kea (568) and Mt. Lemmon Survey (G96). From the
known nominal orbit of this asteroid we obtain
the values
\[
\rho_1 = 1.0419\,,
\hskip 1cm
\rho_2 = 2.0485
\]
of the topocentric distance (in AU) at the two mean epochs of the
observations.

\begin{figure}[ht]
\centerline{\epsfig{figure=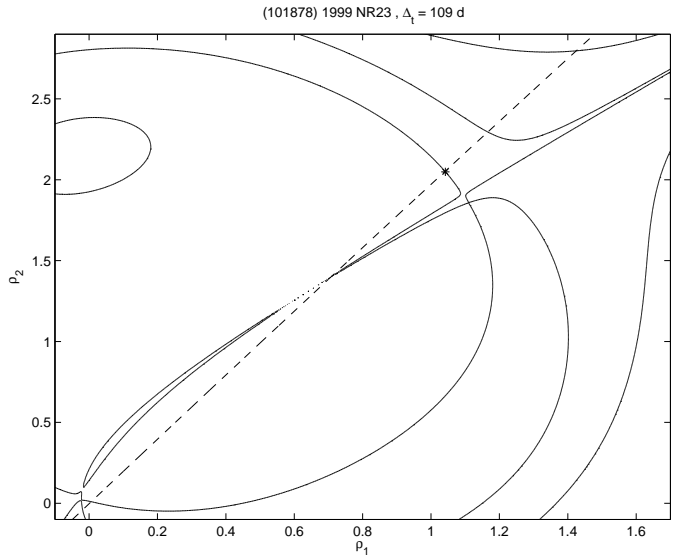,width=9cm}}
\caption{Intersections of the curves $p=0, q=0$ (solid and dashed,
respectively) in the plane $\rho_1,\rho_2$ for the asteroid ($101878$)
$1999$ NR$_{23}$: the asterisk corresponds to the true orbit.}
\label{fig:101878}
\end{figure}

In Figure~\ref{fig:101878} we show the intersections between the
curves defined by $p(\rho_1, \rho_2)$ and $q(\rho_1, \rho_2)$.
By solving the corresponding problem (\ref{intersec}) with the method
described in Subsection~\ref{ss:DFT} we find the 6 positive pairs of
solutions $(\rho_1,\rho_2)$ displayed in Table~\ref{tab:sol_found}.

\begin{table}[ht]
\begin{center}
\begin{tabular}{l|c|c}
       &$\rho_1$         &$\rho_2$\\
\hline
1   &  0.0059   &  0.0097 \\
2   &  0.7130   &  1.4100 \\
3   &  0.7045   &  1.3933 \\
4   &  1.0409   &  2.0517 \\
5   &  1.1659   &  2.2952 \\
6   &  1.4246   &  2.7968 \\
\end{tabular}
\end{center}
\caption{Solutions of the system (\ref{intersec}) for ($101878$)
$1999$ NR$_{23}$.}
\label{tab:sol_found}
\end{table}
\noindent After removing solution 1 (with both components very small),
the spurious solution 6 (not satisfying (\ref{energyeq_fstsquare}))
and the spurious solutions 3 and 5 (not satisfying
(\ref{energyeq_nosquare})), we are left with the values labeled 2 and 4
in Table~\ref{tab:sol_found}.  Note that, even if solutions 2 and 3
look close, they are far apart enough to select only one of them as a
good solution.
\begin{table}[h!]
\begin{center}
\begin{tabular}{c|c|c|}
         &$1$   &$2$\\
\hline
$\Vert\Delta_{1,2}\Vert_\star$   &$487.65806$   &$0.19505$\\
\hline
$a$
&\begin{tabular}{r|r} $6.87384$ &$6.87384$\\ \end{tabular}
&\begin{tabular}{r|r} $2.25828$ &$2.25828$\\ \end{tabular}\\
$e$
&\begin{tabular}{r|r} $0.81798$ &$0.81798$\\ \end{tabular}
&\begin{tabular}{r|r} $0.19787$ &$0.19787$\\ \end{tabular}\\
$I$
&\begin{tabular}{r|r} $0.51733$ &$0.51733$\\ \end{tabular}
&\begin{tabular}{r|r} $0.59995$ &$0.59995$\\ \end{tabular}\\
$\Omega$
&\begin{tabular}{r|r} $156.55521$ &$156.55521$\\ \end{tabular}
&\begin{tabular}{r|r} $156.42531$ &$156.42531$\\ \end{tabular}\\
$\omega$
&\begin{tabular}{r|r} $144.68146$ &$321.78289$\\ \end{tabular}
&\begin{tabular}{r|r} $144.39580$ &$145.26330$\\ \end{tabular}\\
$\ell$
&\begin{tabular}{r|r} $\phantom{00}4.66178$   &$355.27766$\\ \end{tabular}
&\begin{tabular}{r|r} $47.75173$  &$78.65378$\\ \end{tabular}\\
$t$ (MJD)
&\begin{tabular}{r|r} $53999.8205$   &$54109.1368$\\ \end{tabular}
&\begin{tabular}{r|r} $53999.8186$  &$54109.1331$\\ \end{tabular}\\
\end{tabular}
\end{center}
\caption{Keplerian elements (angles in degrees) corresponding to the
  pairs $(\rho_1,\rho_2)$ labeled with 2 and 4 in
  Table~\ref{tab:sol_found}. The value of
  $\Vert\Delta_{1,2}\Vert_\star$ is shown for the two solutions.}
\label{tab:keporb}
\end{table}
For the remaining solutions 2 and 4 we succeed in computing Keplerian
orbits, that we list in Table~\ref{tab:keporb}.
The values of $a, e, I, \Omega$ are the same for each pair: this is
due to the fact that each pair of orbits shares the same angular
momentum and the same energy.  The value of the identification norm
$\Vert\Delta_{1,2}\Vert_\star$, also shown in Table~\ref{tab:keporb},
strongly suggests to select the second solution.  The results appear
pretty good, in fact the differences with the true solution are of the
order of $3\times 10^{-5}$ AU and the errors in the Keplerian elements
are comparable with the planetary perturbations; this is intrinsically
bound to the use of a two-body approximation.

\section{Numerical experiments with simulated observations}
\label{s:simul}

We have tested our identification method with the DFT algorithm,
explained in Subsection~\ref{ss:DFT}, using simulated observations of
objects in a solar system model. The data have been given to us by
R. Jedicke and L. Dennau from the Institute of Astronomy, University of
Hawaii, and the data quality resemble the one which should be achieved
by the Pan-STARRS telescope when it will be fully operative.
The RMS of the observations vary from 0.01 to ~0.02 $arcsec$, that is
rather optimistic for the current surveys.  
%
The current astrometric data quality of the Pan-STARRS 1 telescope is
such that the RMS of the residuals for well determined asteroid orbits
is between 0.11 and 0.13 $arcsec$. Better results should be achieved
when the astrometric reduction of asteroid detections will be
performed with respect to a catalogue generated by the Pan-STARRS
survey itself.

\noindent These observations cover 31 observing nights, in three
consecutive lunations and are grouped into {\em tracklets}.  Each
tracklet is composed by observations presumably belonging to the same
object and covering a short arc: some of them are false (e.g. join
observations of different objects). From each tracklet we can compute
an attributable.
We have first applied to the database of tracklets the identification
procedures defined in \cite{miletal05}, \cite{kubica}. Then we have
tested our method on the leftover database, for which the previous
procedure has failed.  These remaining observations corresponds to
19441 objects, and 24590 tracklets, but only 4132 objects have at
least two tracklets, that is a necessary requirement for the
application of our method.  The hyperbolic orbits have been removed
from the solar system model: in fact our current method does not
search for them, but it could be easily modified to include their
orbit determination.
To each accepted preliminary orbit obtained from a pair of
attributables we apply the differential corrections, using all the
observations at our disposal, to compute a least squares orbit with
its covariance matrix.\footnote{We use the preliminary orbit at time
$\tilde{t}_1$ as starting guess for the differential corrections. We
could also use the orbit at time $\tilde{t}_2$, or an `average orbit'
at time $(\tilde{t}_1+\tilde{t}_2)/2$.}

To reduce the computational complexity, we need to define a filter 
selecting the pairs of attributables which we try to link.  In
Section~\ref{ss:filter} we describe the two filters we have used in
processing the simulated data.

\subsection{Filtering pairs of attributables}
\label{ss:filter}

\subsubsection{First filter: guessing the second angular position.}

A first simple way to discard pairs of attributables at epochs
$\bar{t}_1, \bar{t}_2$ is to constrain the time span $\delta
t=\bar{t}_2-\bar{t}_1$: we require 
\begin{equation}
\delta t_{min} \leq \delta t \leq
\delta t_{max}
\label{deltat_bounds}
\end{equation}
for suitable positive constants $\delta t_{min},\delta t_{max}$.  In
our experiment we have used $\delta t_{max}=99$ days and $\delta
t_{min}=0.5$ days, that practically means we have tried to link
attributables obtained in different nights.
For each given pair of attributables at epochs $\bar{t}_1,
\bar{t}_2$ fulfilling (\ref{deltat_bounds}) we consider for $i=1,2$
the corresponding proper motions $\eta_i$ and the mobile bases $\{
\hrho_i, \hv_i, \hn_i \}$, defined in Section~\ref{s:attrib}.
We want to use one of the proper motions, say $\eta_1$, to
bound the region in the sky where we could recover the object at the
other time $\bar{t}_2$.

\noindent Let us form the orthogonal matrices $V_1=[\hrho_1 | \hv_1 |
\hn_1]$ and $V_2=[\hrho_2 | \hv_2 | \hn_2]$: these are rotation
matrices to the mobile bases $\{ \hrho_i, \hv_i, \hn_i \}$, $i=1,2$.
Let $R_{\phi\, \hat{\bf e}}$ denote the rotation of an angle $\phi$
around the unit vector $\hat{\bf e}$.  Then $R_{\eta_1\delta
t\,\hn_1}=V_1\, R_{\eta_1\delta t\,\hat{\mathbf{z}}}\,V_1^T$
($\hat{\mathbf{z}}$ is the third unit vector of the reference frame
defining our rectangular coordinates) is the parallel transport matrix
along the geodesic on the unit sphere defined by $\Att_1$ to time
$\bar{t}_2$; hence $\hrho_{12} = R_{\eta_1\delta t\,\hn_1} \hrho_1$ is
the predicted observation direction at time $\bar{t}_2$, assuming the
trajectory is a great circle and the proper motion is constant.
By exchanging the order of the two attributables we can compute
$R_{-\eta_2\delta t\,\hn_2} = V_2\, R_{-\eta_2\delta
t\,\hat{\mathbf{z}}}\,V_2^T$ and
$\hrho_{21} = R_{-\eta_2\delta t\,\hn_2}\; \hrho_2$, that is the prediction
at time $\bar{t}_1$.
We use the metric 
\[
d(\hrho_1, \hrho_2) = \min\{ \widehat{\hrho_{12}, \hrho_2},
\widehat{\hrho_{21}, \hrho_1} \}\,,
\]
that is the minimum between the two angular differences, discarding
pairs of attributables that give rise to a large value of this metric. 

Note that the proper motion does not vary too much in the time
interval between the two attributables provided $\delta t_{max}$ is
small enough; thus, if we want to use large values of
$\delta t$, we have also to allow large values of the metric $d$.

\subsubsection{Second filter: symmetric LLS fit.}

Given the two attributables $\Att_1,\Att_2$ at times $\bar{t}_1,\bar{t}_2$ 
we perform a quadratic approximation of the apparent motion on the celestial 
sphere $S^2$ by using a Linear Least Squares (LLS) fit. 
The apparent motion is given by the functions $\alpha(t),\delta(t)$. 
We approximate $\alpha(t),\delta(t)$ with second degree polynomials whose
coefficients are derived from a least squares fit.
We denote the approximating quadratic functions as
\begin{equation}
\alpha(t)=\aq+\daq(t-\bar t)+\frac{1}{2}\ddaq(t-\bar t)^2\,,\hskip 0.8cm
\delta(t)=\delq+\ddelq(t-\bar t)+\frac{1}{2}\dddelq(t-\bar t)^2\,,
\label{model_ad}
\end{equation}
where $\bar t = \frac{1}{2}(\bar{t}_1+\bar{t}_2)$ is the mean of the times 
of the attributables.
The corresponding time derivatives are
\[
\dot\alpha(t)=\daq+\ddaq(t-\bar t)\,,\hskip 0.8cm
\dot\delta(t)=\ddelq+\dddelq(t-\bar t)\,.
\]
We want to determine the 6 quantities $\aq, \daq, \ddaq, \delq,
\ddelq, \dddelq$ using the data coming from the attributables.  
The vector of residuals is
\[
\bxi=
\left(
\Att_1 - \Att(\bar{t}_1)\,, \Att_2 - \Att(\bar{t}_2)\right)^T\,,
\]
with $\Att(t)=(\alpha(t), \delta(t), \alphadot(t), \deltadot(t))$.

\noindent Given the covariance matrices $\Gamma_{\Att_1},
\Gamma_{\Att_2}$ associated to the attributables we use them to {\em
weight} the residuals in the definition of the target function:
\[
Q(\bxi)=\frac{1}{8}\bxi \cdot W \bxi\,,\hskip 0.8cm
\mbox{where}\hskip 0.3cm 
W^{-1}=
\left(
\begin{array}{cc}
\Gamma_{\Att_1} & 0 \\
0 & \Gamma_{\Att_2}
\end{array}
\right)\,.
\]
We introduce the notation
\[
\bx = (\aq,\daq,\ddaq,\delq,\ddelq,\dddelq)^T
\,, \quad
\vec \lambda= (\Att_1,\Att_2)^T\ .
\]
The value of $\bxi=\bxi(\bx)$ that minimizes the target function is
obtained by solving the normal equation
\[
C\bx=-B^T W\vec\lambda\,,\hskip 0.8cm
\mbox{where}\hskip 0.2cm
B=\frac{\partial \bxi}{\partial \bx}, \quad C=B^T WB\,,
\]
and the matrix $B$ has the form
\[
B= -
\left(
\begin{array}{c}
B_1 \\
B_2
\end{array}
\right)
\hskip 0.4cm
\mbox{with}\hskip 0.3cm
B_i=\left(
\begin{array}{cccccc}
1 & (\bar t_i-\bar t) & \frac{1}{2}(\bar t_i-\bar t)^2 & 0 & 0 & 0 \\
0 & 0 & 0 & 1 & (\bar t_i-\bar t) &  \frac{1}{2}(\bar t_i-\bar t)^2\\
0 & 1 & (\bar t_i - \bar t) & 0 & 0 & 0 \\
0 & 0 & 0 & 0 & 1 & (\bar t_i-\bar t) \\
\end{array}
\right)\,,
\]
for $i=1,2$.  Once the value of $\bx$ is given, we compute 
the residuals $\bxi(\bx)$ 
and use the norm $\sqrt{Q(\bxi)}$
to decide which are the pairs of attributables $(\Att_1, \Att_2)$ to
discard. We also discard the pairs giving rise to a large value of the
quantity
\[
\kappa_q\eta_q^2 = \frac{1}{\eta_q} \left[ (\ddot\delta_q \dot\alpha_q
  - \ddot\alpha_q \dot\delta_q) \cos\delta_q + \dot\alpha_q (\eta_q^2
  + {\dot\delta_q}^2)\sin\delta_q \right] \,,
\]
where $\eta_q = \sqrt{\dot\delta_q^2 + \dot\alpha_q^2\cos^2\delta_q}$
and $\kappa_q$ is the geodesic curvature (see \cite{milgro09}, Chapter 9).

\subsection{Results}
\label{ss:results}

The accuracy of the linkage method can be measured by the number of
true identifications over the total number of identifications found.
This computation includes duplications due to alternative solutions.
The total number is $3906$ and the true ones (that may be related to
the same object if it has more than 2 tracklets) are $3144$,
i.e. $80.5\%$ of the total. We could eliminate almost half of the
$762$ false identifications by lowering from 0.15 to 0.06 $arcsec$
the control on the RMS for acceptable orbits after differential
corrections: but this would make us lose 102 true identifications.

In Table~\ref{tab:eff} we show the efficiency of the linkage
procedure, that is we write the number of objects for which at least a
pair of tracklets has been correctly linked, giving the details for
the MB (Main Belt) and the NEO (Near Earth Object) class. As expected, the
efficiency appears greater if there are three tracklets that can be
pairwise linked.
\begin{table}[h!]
\begin{center}
\begin{tabular}{c|c|c|c}
{with 2 tracklets in 2 nights}         &Total   &Found  &Lost\\
\hline
all
&$1074$
&\begin{tabular}{c|c} $963$   &$89.7\%$ \\ \end{tabular}
&\begin{tabular}{c|c} $111$   &$10.3\%$ \\ \end{tabular}\\
MB
&$1038$
&\begin{tabular}{c|c} $947$   &$91.2\%$ \\ \end{tabular}
&\begin{tabular}{c|c} $91$   &$\phantom{0}8.8\%$ \\ \end{tabular}\\
NEO
&$19$
&\begin{tabular}{c|c} $\phantom{00}9$   &$47.4\%$ \\ \end{tabular}
&\begin{tabular}{c|c} $\phantom{0}10$   &$52.6\%$ \\ \end{tabular}\\
\hline
 &&& \\
\hline
{with 3 tracklets in 3 nights}         &Total   &Found  &Lost\\
\hline
All
&$214$
&\begin{tabular}{r|r} $205$   &$95.8\%$ \\ \end{tabular}
&\begin{tabular}{r|r} $\phantom{00}9$   &$\phantom{0}4.2\%$ \\ \end{tabular}\\
MB
&$197$
&\begin{tabular}{r|r} $196$   &$99.5\%$ \\ \end{tabular}
&\begin{tabular}{r|r} $\phantom{00}1$   &$\phantom{0}0.5\%$ \\ \end{tabular}\\
NEO
&$3$
&\begin{tabular}{r|r} $\phantom{00}2$   &$66.7\%$ \\ \end{tabular}
&\begin{tabular}{r|r} $\phantom{00}1$   &$33.3\%$ \\ \end{tabular}\\
\end{tabular}
\end{center}
\caption{Efficiency of the identification procedure.}
\label{tab:eff}
\end{table}
We stress that we have tested our method with data for which the
other available methods in \cite{miletal05}, \cite{kubica} could not
perform the linkage.

\section{Conclusions and future work}

We have investigated an orbit determination method that is based on
the integrals of the Kepler problem and is suitable to be used with
modern data sets of observations. With the recent technologies we can
collect a large number of tracklets in each observing night and it is
difficult even to relate two tracklets of different nights as
belonging to the same observed object. We have defined a linkage
procedure between tracklets, allowing to compute preliminary orbits
with covariance matrices.  An interesting feature that comes out from
our numerical experiments is that this method appears to work also
when the time span between the two attributables is large, hence it
can be used in cases where other linkage methods
fail.  The efficiency and performance of the algorithm explained
in Subsection~\ref{ss:DFT} have been studied with a large scale test.
Therefore this method can be important for two kinds of applications:
1) to recover objects whose orbit could not be computed with either
the classical or the modern known algorithms; 2) to design the
scheduler of new surveys planning a smaller number of observations for
each object.

The number of alternative solutions of the problem 
deserves a deeper investigation, however we expect that
the acceptable ones should often be much less than 48, the total
degree of the polynomial system (\ref{intersec}).  Moreover the
performance of the second algorithm to solve (\ref{intersec}),
described in Subsection~\ref{ss:normal_form}, has not been tested yet:
we would like to perform further experiments to decide if it allows to
decrease the computation time.

\section{Acknowledgements}
We wish to thank Massimo Caboara, from the University of Pisa, for
his useful suggestions on the algebraic aspects of this work. We are
also grateful to Robert Jedicke and Larry Dennau, from the MOPS team
of the Pan-STARRS project, for providing us with the simulation data
used in Section~\ref{s:simul}.


\end{document}